\documentclass[a4paper,12pt,preprint,aps,floatfix,unsortedaddress]{revtex4}
\usepackage{graphicx}
\usepackage{amsmath}
\usepackage{longtable,array}
\usepackage{lscape}
\usepackage{color}
\usepackage{epsfig}

\newcommand{\etal}{{\it et al.} }
\newcommand{\ai}{{\it ab initio}}

\newcommand{\cm}{cm$^{-1}$}

\newcommand{\hato}{H$_2$$^{16}$O}

\def\a0{{$a_{\rm 0}$}}

\begin{document}

\title{ High accuracy water potential energy surface
for the calculation of infrared spectra}

\author{Irina I. Mizus, Aleksandra A. Kyuberis, Nikolai F. Zobov, Vladimir Yu. Makhnev}
\affiliation{Institute of Applied Physics, Russian Academy of Science,
Ulyanov Street 46, Nizhny Novgorod, Russia 603950}

\author{ Oleg L. Polyansky, Jonathan Tennyson }                     
\affiliation{Department of Physics and Astronomy, University College London,
Gower Street, London WC1E 6BT, United Kingdom}

\date{\today}

\begin{abstract}
  Transition intensities for small molecules such as water and CO$_2$
  can now be computed with such high accuracy that they are being used
  to systematically replace measurements in standard databases. These
  calculations use high accuracy \ai\ dipole moment surfaces and
  wavefunctions from spectroscopically-determined potential energy
  surfaces. Here an extra high accuracy potential energy surface
  (PES) of the water molecule (\hato) is produced starting from an
  \ai\ PES which is then refined to empirical rovibrational energy
  levels. Variational nuclear motion calculations using this PES
  reproduce the fitted energy levels with a standard deviation of
  0.011 \cm, approximately three times their stated uncertainty. Use
  of wavefunctions computed with this refined PES is found to 
  improve the predicted transition intensities for selected
  (problematic) transitions. A new room temperature line list for
  \hato\ is presented. It is suggested that the associated set of line
  intensities is the most accurate available to date for this
  species.
\end{abstract}

\maketitle
\section{Introduction} 

Most processes in chemical physics are governed by potential energy
surfaces (PESs) and access to accurate PESs therefore allows the
accurate prediction of properties.  In a series of papers Murrell and
co-workers developed analytic methods of representing PESs for small
(usually triatomic) molecules
\cite{analyt1} -- \cite{analyt7}.
Surfaces predicted on the basis of \ai\ electronic structure
calculations were usually improved by comparison with or fitting to
spectroscopic data. This is precisely the technique we employ here.
Much of this work on PESs is captured in the book ``Molecular
Potential Energy Functions'' by Murrell \etal\ \cite{PESbook.Murrell}.
                              
Water molecule is the number one molecule in the HITRAN database
\cite{jt691}, which reflects its importance in the Earth's
atmosphere, and is a key constituent of other solar system bodies,
exoplanets and cool stars \cite{09Bexxxx.exo}.  It is therefore
unsurprising that Sorbie and Murrell, in their seminal paper on
constructing PESs, chose to concentrate on water \cite{analyt1}. Many
other groups have subsequently followed their lead in constructing
accurate semi-empirical PESs for water
\cite{81MuCaMiGu,ch87,hc88,jt149,ps97,jt423}, the current authors
included \cite{jt150,jt182,jt250,jt375,11BuPoZo.H2O,jt519}.  An
important aspect of the study of the water molecule and the use of the
results of these studies in the modelling of different terrestrial and
astrophysical environments is the ability to accurately predict line
intensities.

Over the last two decades, significant improvements have been made the
accuracy of the water PESs; residuals in predicted rotation-vibration
line positions have dropped from about 0.6 \cm \cite{jt150,jt182} to
0.025 \cm\ \cite{ps97,11BuPoZo.H2O}. However, still this figure is
about an order-of-magnitude larger than the uncertainty in empirical
determinations of rotation-vibration energy levels \cite{jt275,jt539}.
Improving these predictions towards experimental accuracy places high
significance on every step of the calculation. As demonstrated below,
these improvements are important not only for accurately reproducing
line positions, but also for the generation of accurate wavefunctions
for use in intensity calculations.

A testimony to the improvement in the accuracy of computed infrared
transition intensities based on the use of \ai\ dipole moment surfaces
(DMSs) is the systematic adoption of computed intensities in the place
of measured one for both water and CO$_2$ in the recent (2016) release of
HITRAN \cite{jt691}. Use of this methodology is particularly powerful
for isotopically substituted species for which the accuracy of the
computations is essentially unchanged \cite{jt667,jt678,jt690}, but
experimental determinations become much harder.  These studies employ
the Lodi-Tennyson method \cite{jt522} which uses stability analysis
based on calculations using two different PESs and two different DMSs
to identify ``unstable'' transitions for which the computed intensities are
not reliable. Significant differences between intensities calculated
using different PESs for certain transitions have been found for water
\cite{jt687,jt690}, CO$_2$ \cite{jt625} and H$_2$CO \cite{jt597}.
Conversely, for many transitions it has proved possible to compute
the line intensities for water \cite{jt509,jt687} and CO$_2$
\cite{jt613} with sub-percent accuracy, which reflects the accuracy of
the underlying \ai\ DMS.  To make further progress on this problem it
becomes important to estimate and eliminate the causes of the
remaining uncertainties.  The largest of these appears to be the
calculation of wavefunctions for states which are affected by
accidental interactions with other states of the same overall
symmetry; spectroscopers generally call these interactions resonances. It
is the failure to precisely treat these resonances that leads to
the unstable transitions. As shown below, better treatment can be achieved by
systematic improvement of the PES and hence the wavefunctions.

Section II describes the procedure used to obtain our new PES. 
Section III presents the computed energy levels obtained using 
our new PES. Section IV compares and discusses calculations
of intensities using different PESs.
Section V describes our new, room temperature \hato\ line list.
Section VI presents our conclusions and plans for further work.

\section{New potential energy surface}
As a starting point for the optimization process we used a semi-empirical PES 
due to Bubukina \emph{et al.} \cite{11BuPoZo.H2O}. Bubukina \emph{et al.}  in turn used
the high quality CVRQD \ai\ PES \cite{jt309,jt394} as their starting point and
augmented this surface with corrections for
adiabatic \cite{jt180}, relativistic  \cite{jt274}, and quantum electrodynamics (QED) 
\cite{jt265} effects. This fully \ai\ starting point predicts rotation-vibration
energy levels with an accuracy of about 1 \cm.

\subsection{Nuclear motion calculation}\label{fit_pars}
Variational nuclear motion calculations were performed in Radau
coordinates using DVR3D \cite{jt338}. Morse-like oscillators with the
values of parameters $r_e = 2.55$, $D_e = 0.25$ and $\omega_e = 0.007$
in atomic units were used for both radial coordinates, and associated
Legendre functions for the angular coordinate as basis functions. The
corresponding discrete variable representation (DVR) grids contained
29, 29 and 40 points for these coordinates, respectively. The final
diagonalized vibrational matrices had a dimension of 1500. For the
rotational problem, the dimensions of final matrices can be obtained
as $400(J + 1 - p)$, where $J$ is the total angular momentum quantum
number and $p$ is the value of parity. Only nuclear masses were used
in these calculations.  To ensure good accuracy for high-$J$
calculations, we followed Bubukina \emph{et al.} \cite{11BuPoZo.H2O}'s
approach to rotational non-adiabatic effects which adopts a
simplification \cite{jt282} of \ai\ procedure developed by Schwenke
\cite{03Schwen.method} for treating non-Born-Oppenhemer effects. The
values of the adjustable parameters used by Bubukina \emph{et al.} to
scale Schwenke's results were left unchanged.

\subsection{Optimization results}\label{res}
The optimization procedure was based on a method developed by Yurchenko 
\emph{et al.} \cite{03YuCaJe.PH3}.

This method allows the fit to simultaneously optimize reproduction of 
empirical energy levels and \emph{ab initio} grid points. This procedure
helps the fit to avoid nonphysical behavior in the optimized PES. For
this purpose we used a set of 677 \emph{ab initio} energies computed
by Grechko \emph{et al.} \cite{jt467} in the energy region up to
$25\,000\,\text{cm}^{-1}$ (about 2.7\% of \emph{ab initio} points from
the original set of 696 energies were excluded from the fit).

In the fit, we varied the values of 240 potential parameters of the
starting PES to obtain the best predicts of the most-accurate
available empirical energy levels \cite{jt539}. As a result, we obtain
a potential which reproduces the set of 847 empirical energy levels
with $J$ values 0, 2 and 5, and lying below $15\,000\,\text{cm}^{-1}$
with a standard deviation of 0.011$\,\text{cm}^{-1}$.  The initial fit
had the \emph{ab initio} grid points weighted at $10^{-4}$ the
empirical data to ensure that PES remained physical, i.e. did not
develop holes, in the region of interest. For the final stages of the
fit this weighting was reduced to $10^{-8}$.  About 3\% of energy
levels from the complete empirical set in the energy region of
interest (26 from the total set of 873 energies) were excluded from
the fit. Most of these levels (about 20) have high values of bending
quantum number $\nu_2$ and sample a region of the potential which is
not well-characterized by fit.  The standard deviation of our final
PES with respect to the set of Born-Oppenheimer \emph{ab initio}
points is about 68.8$\,\text{cm}^{-1}$, a figure which depends
strongly on high energy points.

Because the PES is designed for studies of states up to $15\,000\,\text{cm}^{-1}$ we refer to
it as PES15k below. A Fortran program giving the PES15k potential is given in the
supplementary material.

\section{Results of the energy levels calculations}
Table I presents  results of a $J=0$ calculations
using PES15k.

\begin{table}
\caption{Observed vibrational band origins below $15\,000\,\text{cm}^{-1}$ calculated with the new PES; values are in \cm.}
     \begin{tabular}{cccrccccr}
\hline\hline
 $v_1v_2v_3$         &        obs     &     calc     &     obs-calc   &~~~& $v_1v_2v_3$&        obs     &     calc     &     obs-calc  \\
\hline                                                                
 0  1  0  &      1594.7463 &   1594.7523  &     0.0060    &&   0  0  1  &      3755.9285 &   3755.9277  &    -0.0009    \\
 0  2  0  &      3151.6298 &   3151.6453  &     0.0155    &&   0  1  1  &      5331.2673 &   5331.2658  &    -0.0015    \\
 1  0  0  &      3657.0533 &   3657.0406  &    -0.0126    &&   0  2  1  &      6871.5202 &   6871.5225  &     0.0023    \\
 0  3  0  &      4666.7905 &   4666.7948  &     0.0043    &&   1  0  1  &      7249.8169 &   7249.8175  &     0.0006    \\
 1  1  0  &      5234.9756 &   5234.9759  &     0.0004    &&   0  3  1  &      8373.8514 &   8373.8529  &     0.0015    \\
 0  4  0  &      6134.0150 &   6134.0207  &     0.0057    &&   1  1  1  &      8806.9990 &   8806.9972  &    -0.0018    \\
 1  2  0  &      6775.0935 &   6775.0906  &    -0.0029    &&   0  4  1  &      9833.5829 &   9833.5805  &    -0.0025   \\
 2  0  0  &      7201.5399 &   7201.5414  &     0.0016    &&   1  2  1  &     10328.7293 &  10328.7248  &    -0.0044    \\
 0  0  2  &      7445.0562 &   7445.0269  &    -0.0293   &&   2  0  1  &     10613.3563 &  10613.3510  &    -0.0053    \\
 0  5  0  &      7542.3725 &   7542.4207  &     0.0482    &&   0  0  3  &     11032.4041 &  11032.3939  &    -0.0102    \\ 
 1  3  0  &      8273.9757 &   8273.9713  &    -0.0044    &&   0  5  1  &     11242.7757 &  11242.7716  &    -0.0041    \\
 2  1  0  &      8761.5816 &   8761.5851  &     0.0035    &&   1  3  1  &     11813.2069 &  11813.2046  &    -0.0023    \\
 0  1  2  &      9000.1360 &   9000.1258  &    -0.0103    &&   2  1  1  &     12151.2539 &  12151.2438  &    -0.0101    \\
 2  2  0  &     10284.3644 &  10284.3588  &    -0.0055    &&   0  1  3  &     12565.0064 &  12564.9970  &    -0.0094    \\
 0  2  2  &     10521.7577 &  10521.7435  &    -0.0142    &&   1  4  1  &     13256.1550 &  13256.1584  &     0.0034    \\
 3  0  0  &     10599.6860 &  10599.6789  &    -0.0071    &&   2  2  1  &     13652.6532 &  13652.6523  &    -0.0009    \\
 1  0  2  &     10868.8747 &  10868.8608  &    -0.0139    &&   3  0  1  &     13830.9368 &  13830.9336  &    -0.0032    \\
 2  3  0  &     11767.3890 &  11767.3673  &    -0.0217    &&   0  2  3  &     14066.1936 &  14066.1769  &    -0.0167    \\
 0  3  2  &     12007.7743 &  12007.7778  &     0.0034    &&   1  0  3  &     14318.8121 &  14318.8008  &    -0.0113    \\
 3  1  0  &     12139.3153 &  12139.3065  &    -0.0088    &&   1  5  1  &     14647.9733 &  14647.9996  &     0.0263    \\
 1  1  2  &     12407.6620 &  12407.6465  &    -0.0155    \\
 3  2  0  &     13640.7166 &  13640.6811  &    -0.0355    \\
 4  0  0  &     13828.2747 &  13828.2615  &    -0.0132    \\
 1  2  2  &     13910.8936 &  13910.8796  &    -0.0140    \\
 2  0  2  &     14221.1585 &  14221.1459  &    -0.0126    \\
 0  0  4  &     14537.5043 &  14537.4859  &    -0.0184    \\
\hline\hline
        \end{tabular}
 \end{table}

 Table II presents  results of a comparison of the standard
 deviations for states with rotational quantum numbers $J$ up to 15
 and lying below $15\,000\,\text{cm}^{-1}$ using the PES of Bubukina
 \emph{et al.} \cite{11BuPoZo.H2O} and PES15k.  These standard
 deviations are computed from the discrepancies between calculated and
 experimental levels as follows. Empirical energy levels up to $J =
 15$ were taken from the recent IUPAC compilation \cite{jt539,jt562}.
 Not all the levels were included for the comparison. Some of the
 levels - about 2\% of those derived in \cite{jt539} were excluded as
 they were outliers. There are two reasons that certain levels might
 show larger discrepancies.  First, it could be the genuine
 experimental inaccuracy or incorrectly incorporated data. A number of
 problems with the IUPAC compilation of energy levels have been
 identified and are in the processes of being fixed
 \cite{jtwaterupdate}.  Second, our model becomes worse for very
 highly excited $v_2$ bending vibrational quantum numbers and
 correspondingly for very highly excited $K_a$ rotational quantum
 numbers. There are particular issues with quantum numbers of high
  $v_2$ states \cite{jt234} and transitions to these states
 are in general much weaker than the other transitions, thus the
 influence of such lines on the accuracy of any line list is minor.
 As we can see from Table II, the improved average accuracy, expressed
 in the form of the standard deviation, varies between a factor of 1.5 to 2,
 which is significant.

\begin{table}
  \caption{Standard deviation, in \cm, as function
of rotational excitation, $J$, of  \hato\ energy levels 
obtained from calculations with two PESs with respect to the
empirical energy levels \cite{jt539}; $N$ gives the number
of levels considered for each $J$.} 
     \begin{tabular}{cccc}
\hline\hline
 J &   N   &Bubukina {\it et al} \cite{11BuPoZo.H2O}&    PES15k  \\
\hline
 0 &   41  &    0.0145    &    0.0108     \\
 1 &  148  &    0.0181    &    0.0116     \\
 2 &  249  &    0.0239    &    0.0110     \\
 3 &  356  &    0.0186    &    0.0109     \\
 4 &  450  &    0.0166    &    0.0101     \\
 5 &  548  &    0.0174    &    0.0092     \\
 6 &  623  &    0.0169    &    0.0104     \\
 7 &  663  &    0.0162    &    0.0111     \\
 8 &  682  &    0.0238    &    0.0153     \\
 9 &  662  &    0.0257    &    0.0188     \\
10 &  610  &    0.0280    &    0.0215     \\
11 &  586  &    0.0328    &    0.0260     \\
12 &  545  &    0.0400    &    0.0288     \\
13 &  495  &    0.0432    &    0.0313     \\
14 &  441  &    0.0483    &    0.0335     \\
15 &  387  &    0.0491    &    0.0347     \\
\hline
\hline
        \end{tabular}
 \end{table}

\section{Comparison of intensities}

The intensity of a transition depends on the square of the transition dipole.
The transition dipole between two states can be computed using the expression:
\begin{equation}
\mu_{if} = \sum_t < i | \mu_t | f >
\end{equation}
where for a vibration-rotation transition, the initial and final
states are represented by nuclear motion wave functions $|i>$ and
$|f>$, and the sum runs over the components of the internal dipole
moment vector, $\underline{\mu}$.  For a given DMS, differences in the values of the
intensities reflect differences in the wavefunctions obtained as a
result of the solution of the nuclear-motion Schr\"odinger equation
for a certain PES. In this work all calculations use the LTP2011 DMS
of Lodi {\it et al} \cite{jt509}, which is the most accurate
DMS currently available.

The steady improvement in intensity measurements
\cite{jt432,11WuViJo.CO2,jt613,17GhLiFl.CO2,17SiHoxx.H2O,jt700,17SiHoxx.H2O}
towards the sub-percent level of accuracy has paralleled attempts to
improve the accuracy of intensity calculations.  Given an accurate,
\ai\ DMS \cite{jt509,jt613}, the Lodi-Tennyson method \cite{jt522} tests for
unstable transitions, whose intensities can not be determined
accurately. These transitions are identified on the basis of four line
list calculations using two PESs and two DMSs. The usefulness of the
Lodi-Tennyson method depends strongly on the accuracy of both the
primary PES and the secondary PES.  For example, attempts to use an
\ai\ PES as a secondary PES normally results in a significant
overestimation of number of unstable lines \cite{jt625}. Thus, we need
extremely accurate PESs for at least three reasons. The first is
determination of the most accurate line positions in the calculation
of molecular line lists. The second is for accurate wavefunctions which
can be used for the accurate intensity calculations. The third is that an
excellent pair of PESs are necessary for the Lodi-Tennyson-style
evaluation of stability.

 For these purposes it is important to evaluate what is the difference in intensity
accuracy for the different lines using the same DMS and two different 
very accurate PESes. As it is shown in the previous section,
the accuracy of the PES of this work is extremely high, whereas the
accuracy of the PES of Bubukina {\it et al} \cite{11BuPoZo.H2O} is also very high, it becomes
an interesting problem to compare the intensities coming from these two PESes.
For this purpose we computed linelist  considering all transitions below 15~000 \cm\ involving
states with up to $J = 4$
using these two PES.  Table III presents a comparison of the differences in intensities of resulting 8701 lines. 
It can be seen that for most of the lines, more than 75~\%, the transition intensity
is essentially unchanged (changes by less than 0.1\%) by the change wavefunctions, while approximately
1\%\ of the lines change by more than 5\%.

\begin{table}
\caption{Comparison of intensities computed using wavefunctions
from the PES15k and Bubukina {\it et al} \cite{11BuPoZo.H2O}.
The comparison is for 8686 transitions
between states with $J \leq 4$. The percentage differences
are computed as $100\times$($I$(PES15)-$I$(Bubukina))/$I$(PES15).}
     \begin{tabular}{lr}
\hline
\hline
\%  &number of lines  \\
\hline
$\geq$ 20\%   &   8     \\
5\%-20\%      &   73      \\
2\%-5\%       &  213      \\
1\%-2\%       &  292      \\
0.2\% - 0.5\% & 1090      \\
   $\leq$ 0.1\% & 4974 \\                       
\hline\hline          
        \end{tabular}     
 \end{table}  
     
           Table IV presents an example of the calculation of intensities of
water absorption lines in the small spectra region
between 3495 and 3500 \cm\ using wavefunctions calculated by two
different PESs: PES15k (this work) and Bubukina {\it et al} \cite{11BuPoZo.H2O}.
For many lines, the difference is less than 0.1 \%.  Some of the
lines intensities differ more significantly by around 0.2 \%. There are also a few lines with
a significant difference in intensities of more than 1\%. For one line the difference is 13 \%;
this line was previously deemed unstable using the Lodi-Tennyson method \cite{jt687}.

\begin{table}
\caption{Sample comparison of intensity calculations for the  region
3495 -- 3500 \cm. Experimental measurements are taken from  Loos \etal\ \cite{17LoBiWa,jt687}
and have uncertainties better than 1\%. The predicted intensities used
wavefunctions calculated using 
PES15k (this work) and the PES of Bubukina {\it et al} \cite{11BuPoZo.H2O}. }
     \begin{tabular}{lccccrrrrr}
\hline \hline
Frequency& \multicolumn{4}{c}{Assignment} & $I$(exp) & $I$(PES15k) & $I$(Bubukina) & $\delta I$(PES15k)/\%& $\delta I$(Bubukina)/\% \\ 
\hline
3495.065 & 1  0  0 & 13  3 10  &     0  0  0 &13  4  9 & 2.62($-$25) & 2.6063($-$25) & 2.6054($-$25) &$-$0.45 &$-$0.48  \\
3495.176 & 1  0  0 &  6  2  5  &     0  0  0 & 7  1  6 & 1.62($-$21) & 1.6070($-$21) & 1.6086($-$21) &$-$0.93 &$-$0.83 \\
3495.506 & 1  0  0 & 13  5  8  &     0  0  0 &14  4 11 & 1.44($-$25) & 1.4901($-$25) & 1.4904($-$25) & 3.36  & 3.38  \\
3496.164 & 1  1  0 &  2  1  2  &     0  1  0 & 3  2  1 & 1.51($-$24) & 1.4784($-$24) & 1.4776($-$24) &$-$2.07 &$-$2.12  \\
3496.279 & 1  0  0 &  9  4  6  &     0  0  0 & 9  5  5 & 9.76($-$24) & 9.6424($-$24) & 9.6300($-$24) &$-$1.25 &$-$1.38  \\
3496.383 & 1  0  0 & 11  5  6  &     0  0  0 &12  4  9 & 6.21($-$24) & 6.1797($-$24) & 6.1846($-$24) &$-$0.54 &$-$0.46 \\
3496.624 & 1  0  0 &  5  1  4  &     0  0  0 & 6  2  5 & 3.30($-$21) & 3.2581($-$21) & 3.2555($-$21) &$-$1.28 &$-$1.37  \\
3496.917 & 1  0  0 & 11  7  5  &     0  0  0 &12  6  6 & 3.73($-$25) & 3.7945($-$25) & 3.8019($-$25) & 1.75  & 1.94  \\
3497.601 & 0  2  0 &  5  5  0  &     0  0  0 & 6  2  5 & 2.29($-$23) & 2.1521($-$23) & 2.4768($-$23) &  $-$6.27 & 7.66   \\
3497.985 & 1  0  0 & 10  6  5  &     0  0  0 &11  5  6 & 1.29($-$23) & 1.2870($-$23) & 1.2898($-$23) &   0.00  & 0.22   \\
3498.602 & 0  0  1 &  6  3  4  &     0  0  0 & 6  5  1 & 6.70($-$23) & 6.6933($-$23) & 6.7045($-$23) &  $-$0.06 & 0.11  \\
3499.420 & 1  1  0 &  6  3  4  &     0  1  0 & 6  4  3 & 1.58($-$25) & 1.5306($-$25) & 1.5297($-$25) &  $-$3.03 &$-$3.09 \\
3499.561 & 0  0  1 & 15  3 13  &     0  0  0 &15  3 12 & 7.71($-$26) & 7.5541($-$26) & 7.5730($-$26) &  $-$2.00 &$-$1.74  \\
3499.746 & 0  0  1 &  8  5  3  &     0  0  0 & 9  5  4 & 7.79($-$22) & 7.8226($-$22) & 7.8306($-$22) &   0.47  & 0.57  \\
3499.925 & 0  2  0 &  7  3  4  &     0  0  0 & 6  2  5 & 8.07($-$23) & 8.1491($-$23) & 8.1537($-$23) &   0.99  & 1.05  \\
\hline  \hline                 
        \end{tabular}    
 \end{table}

 Figure~1 gives a similar comparison with the recent high accuracy
 measurements of Sironneau and Hodges \cite{17SiHoxx.H2O} who consider
 70 unblended water lines in the transparency window region from 7710
 to 7920 \cm; the stated uncertainty of these measurements is only
 0.20\%. For most transitions the difference between the predictions
 of calculations performed with wavefunctions from the PES15k and
 Bubukina {\it et al.} PESs is small. However, in general our new
 calculations with PES15k give results closer to the measurements.

\begin{figure}
\centering
\scalebox{0.5}{\includegraphics{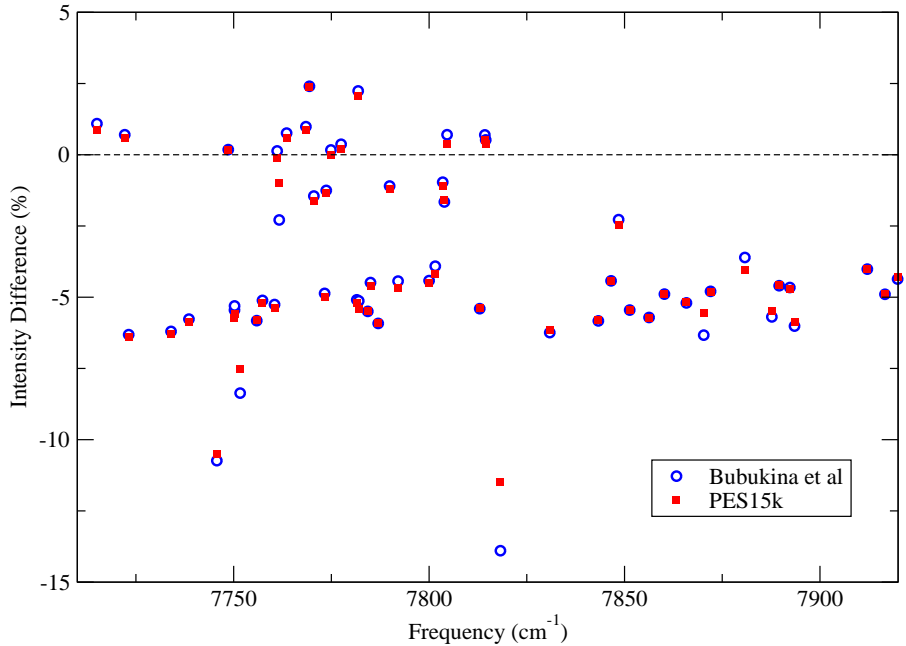}}
\caption{Differences, percent of the observed value, for transition intensities predicted using wavefunctions from the PES15k and Bubukina {\it et al.}  \cite{11BuPoZo.H2O}. The observed data is taken from the recent high accuracy
measurements by Sironneau and Hodges \cite{17SiHoxx.H2O}.}
\label{f:1}
\end{figure}

\section{Line list calculation   }       

Given the improved quality of the intensities generated with the new
PES15k potential, we have computed a new room temperature line list for
\hato. The line list uses the PES15k PES and the LTP2011 DMS
\cite{jt509}. The line list includes all transitions with an intensity
greater than $10^{29}$ cm/molecule at 296~K involving states with $J
\leq 15$ and transitions wavenumbers below 18~000 \cm. This line list,
which follows the HITRAN convention which scales the intensities to
natural abundance, contains 132~034 transitions.  It is given in
HITRAN format as part of the supplementary data.

\section{Conclusions              }                     
 
Since the pioneering work of Murrell and his group on the
representation of potential energy surfaces for polyatomic molecules,the
story has been one of steady improvement. The best surfaces are now
capable of giving predictions competitive with spectroscopic
measurements \cite{jt512}. Of course a PES only exists within the
Born-Oppenheimer approximation so the generation of high accuracy
potentials necessary brings beyond Born-Oppenheimer effects into play.
How best to include these is a subject of active research for small
molecules such as water
\cite{03Schwen.method,jt566,17ScAgSe}.

There has been much recent emphasis on using highly accurate PES and
{\it ab initio} DMS to compute transition intensities with accuracies
approaching that achieved by many experimental studies.
Understandably, much emphasis has been placed on \ai\ techniques for
computing high accuracy DMS \cite{jt509,jt424,sp00}. However, the PES used
plays an important role in providing reliable wavefunctions; this is
particularly important for those states which are sensitive to
interactions with other nearby states of the same overall $J$ and
symmetry. In this work we provide a new PES of improved accuracy which
is used to provide a new \hato\ line list.  We show that while for the
majority of transitions use of  wavefunctions generated employing this new
PES simply replicate results already available; for a minority of
transitions the use of the new PES gives results which are
significantly different and which generally represent an improvement.

Our analysis shows that for more than 10\%\ of the lines use of our
improved PES results in a change in the predicted intensity by more than
0.1 \%. Given that the most recent studies measure intensities for a
few, carefully chosen \hato\ lines with an accuracy of 0.2\%\
\cite{17SiHoxx.H2O}, improvements of this magnitude are important for
the matching theoretical model.  However, as is clear from the results
presented above and from other published studies \cite{jt687,jt645},
the best currently available \ai\ DMS is not able to provide this
level of for many bands. Work on improving the accuracy and extending
the range of DMS is currently in progress.

\section*{Acknowledgement}

We thank Sergey Yurchenko for helpful discussions during the course of this work,
This work was supported by the UK Natural Environment
Research Council through  grant NE/N001508/1 and the Russian Fund for Fundamental Studies.

\bibliographystyle{apsrev}

\end{document}